\documentstyle[11pt,aaspp4]{article}
\def\etal{\hbox{et~al.}}
\def\ie{\hbox{i.e.}}
\def\eg{\hbox{e.g.}}
\def\etc{\hbox{etc.}}
\def\cf{\hbox{cf.}}
\def\Mpc{\hbox{\,h$^{-1}$\,Mpc}}

\def\kms{\hbox{\,km\,s$^{-1}$}}
\def\Dnsigma{\hbox{$D_n$--$\sigma$}}

\def\beqn{\begin{equation}}
\def\eeqn{\end{equation}}

\begin{document}

\slugcomment{\large \it DRAFT -- 16 February 1996}

\title{The Peculiar Motions of Elliptical Galaxies in Two Distant Regions. \\ 
I. Cluster and Galaxy Selection}

\author{Gary Wegner}
\affil{Department of Physics and Astronomy, 6127 Wilder Laboratory,
Dartmouth College, \\ Hanover, NH 03755-3528}

\author{Matthew Colless}
\affil{Mount Stromlo and Siding Spring Observatories, The Australian 
National University, \\ Weston Creek, ACT 2611, Australia}

\author{Glenn Baggley and Roger L. Davies}
\affil{Department of Physics, South Road, Durham DH1 3LE, United 
Kingdom}

\author{Edmund Bertschinger}
\affil{MIT 6-207, Department of Physics, Cambridge, MA 02139}

\author{David Burstein}
\affil{Department of Physics and Astronomy, Box 871054, Arizona State
University, \\ Tempe, AZ 85287-1504}

\author{Robert K. McMahan, Jr}
\affil{Dept of Physics and Astronomy, University of North Carolina, \\
CB 3255 Phillips Hall, Chapel Hill, NC 27599-3255}
\authoraddr{P.O. Box 14026, McMahan Research Laboratories, 79 Alexander
Drive, Research Triangle, NC 27709}

\author{R. P. Saglia}
\affil{Universit\"{a}ts-Sternwarte M\"{u}nchen, Scheinerstra{\ss}e 1, 
D-81679 M\"{u}nchen, Germany}

\begin{abstract}
The EFAR project is a study of 736 candidate elliptical galaxies in 84
clusters lying in two regions towards Hercules-Corona Borealis and
Perseus-Pisces-Cetus at distances $cz$$\approx$6000--15000\kms. In
this paper (the first of a series) we present an introduction to the
EFAR project and describe in detail the selection of the clusters and
galaxies in our sample. Fundamental data for the galaxies and clusters
are given, including accurate new positions for each galaxy and
redshifts for each cluster. The galaxy selection functions are
determined using diameters measured from Schmidt sky survey images for
2185 galaxies in the cluster fields. Future papers in this series will
present the spectroscopic and photometric observations of this sample,
investigate the properties of the fundamental plane for ellipticals,
and determine the large-scale peculiar velocity fields in these two
regions of the universe.
\end{abstract}


\section{Introduction}

This paper is the first in a series reporting the results of the EFAR
project studying the properties and peculiar motions of elliptical
galaxies and clusters in two volumes of the universe at distances
between 6000 and 15000\kms. The aims of this extensive observational
program are: (i)~to study the intrinsic properties of elliptical
galaxies in clusters by compiling a large and homogeneous sample with
high-quality photometric and spectroscopic data; (ii)~to test possible
systematic errors, such as environmental dependence, in existing
elliptical galaxy distance estimators; (iii)~to seek improved distance
estimators based on a more comprehensive understanding of the
properties of ellipticals and how these are affected by the cluster
environment; and (iv)~to determine the peculiar velocity field in
regions that are dynamically independent of the mass concentrations
within 6000\kms in order to test whether the large-amplitude coherent
flows seen locally are typical of bulk motions in the universe.

The EFAR project was conceived in 1986 as a natural progression from
the work of the Seven Samurai (7S: Dressler \etal\ 1987a; Lynden-Bell
\etal\ 1988), who studied the peculiar velocity field traced by
elliptical galaxies closer than 6000\kms. The major finding of that
work was that the local region of the universe was dominated by
large-scale, large-amplitude coherent motions. This result has been
substantially confirmed both by further analysis (Faber \& Burstein
1988; Bertschinger \etal\ 1990) and by subsequent observational
studies, mostly employing the independent Tully-Fisher distance
estimator for spiral galaxies (Aaronson \etal\ 1986, 1989; Han \&
Mould 1990; Willick 1990, 1991; Mathewson \etal\ 1992; Mould \etal\
1993; Courteau \etal\ 1993; Mathewson \& Ford 1994). Although something is
known about the peculiar velocity field within 6000\kms\, 
the nature of the mass concentrations causing the flow
remains controversial (see reviews by Bertschinger 1990; Burstein
1990; Dekel 1994; Strauss \& Willick 1995) and relatively little is known
about galaxy motions far away.

Initial comparisons of the velocity field with the
predictions of cosmological models (Vittorio \etal\ 1986; Bertschinger
\& Juszkiewicz 1988) suggested that the observed motions were
difficult to reconcile
with the favoured biased CDM models that they considered. 
The immediate question raised was
whether the local volume was indeed typical of other regions of the
universe (implying that the standard cosmological models were
incorrect) or whether the local motions are merely an unusual
statistical fluctuation in the universal velocity field. More recent
analyses (Kaiser 1988, 1991; Feldman \& Watkins 1994; Seljak \&
Bertschinger 1994; Dekel \etal\
1996), suggest that the local motions are consistent with the
COBE-normalised standard CDM model. Nonetheless, whether or not the 
local motions are typical of the universe at large remains an
important question. In order to answer this we are measuring 
the peculiar motions in similarly-large regions at sufficient distances 
to be dynamically independent of the local volume studied by 7S and 
most other workers.

During the period it has taken to complete the EFAR observing program,
however, a wider variety of questions have arisen. Chief among
these has been a more searching inquiry as to the reliability of the
$\Dnsigma$ distance indicator developed by 7S (Dressler \etal\ 1987b;
Lynden-Bell \etal\ 1988) and the physical origin of the fundamental
plane (FP) of elliptical galaxies (Djorgovski \& Davis 1987; 
Bender \etal\ 1992; Saglia \etal\ 1993a; 
J{\o}rgensen \etal\ 1993; Pahre \etal\ 1995) of which it
is simply a convenient projection.  Various authors have suggested
that there may be variations in the FP which correlate with galaxy
environment, either directly, through mechanisms such as tidal
stripping (Silk 1989), or indirectly via different stellar populations
(Gregg 1992, 1995). Such effects could lead to significant systematic
errors in any distance estimators based on the FP relations.  Some
claims have been made for the detection of such variations (de
Carvalho \& Djorgovski 1992; Guzman \& Lucey 1993). Other studies,
however, find little variation of the FP with environment (Lynden-Bell
\etal\ 1988; Burstein \etal\ 1990; Lucey \etal\ 1991), and comparisons
of FP and $\Dnsigma$ distance estimates with those derived from
relatively independent and perhaps more accurate estimators, such as
the Tully-Fisher and surface brightness fluctuation methods, show
good agreement (Jacoby \etal\ 1992).

These concerns, and the focus they bring on the formation and
evolution of the elliptical galaxy population, have become as
important a motivation for this work as the original goal of measuring
the peculiar velocity field at large distances. The EFAR project's
goal of measuring the peculiar motions of {\em distant} ellipticals
from a {\em large} and {\em homogeneous} sample, provides a test of
the FP distance estimators that is both {\em severe} (since systematic
errors in peculiar velocities 
are amplified at large distances) and {\em fair} (given the
difficulties of comparing the FP for differently selected samples
observed in different studies). It is worth noting that the 7S dataset
is still the largest in the literature for elliptical galaxies, 
though it was obtained a
decade ago and is based on photoelectric aperture photometry and IDS
spectroscopy. The EFAR sample is comparable in size to that of 7S and
is based largely on CCD imaging and spectroscopy, which confer a
number of advantages in the attempt to reduce observational errors.

Even if {\em systematic} errors prove negligible, the relatively large
($\sim$20\%) {\em random} errors in the $\Dnsigma$ and Tully-Fisher
distance estimators limit exploration of the velocity field beyond
about 6000\kms\ (the `local' region). The only studies which have
attempted to measure the velocity field as far out as 15000\kms\ are
those of Lauer \& Postman (1994), using brightest cluster galaxies as
distance estimators, and Riess \etal\ (1995), using Type Ia
supernovae. These sparse, all-sky samples are suitable for measuring
the convergence of the Local Group dipole motion to the cosmic
microwave background dipole (or lack thereof), but they do not probe
the velocity field on scales of tens of Mpc. This requires the use of
distance estimators which have greater precision per object and/or
apply to clustered objects, allowing denser sampling.

FP-based distance estimators for elliptical galaxies fulfill these
criteria. As recent work by J{\o}rgensen \etal\ (1993) has shown, the
FP can yield distances with errors as low as 11\% per galaxy
(compared to 17\% for $\Dnsigma$ distances based on the same data, and
25\% for the original $\Dnsigma$ distances obtained by the 7S). For
individual {\em clusters} it is possible to reduce the distance errors
by $\surd N$, where $N$ is the number of galaxies in each
cluster. With 5--15 ellipticals per cluster it is therefore possible in
principle to measure individual cluster distances with 3--5\% precision,
corresponding to peculiar velocity errors of 500-750\kms\ at a
distance of 15000\kms. Thus with sufficient clusters in a given volume
it becomes feasible to measure the peculiar velocity field with enough
precision to reliably detect large-scale coherent motions at distances
out to 15000\kms which have amplitudes comparable to those observed in
the local volume within 6000\kms. Provided significant sources of
systematic error in the FP distance estimator can be ruled out or
corrected for, the potential exists to determine the peculiar
velocity fields in distant regions and thus further constrain
cosmological models.

Two preparatory papers have discussed the photoelectric photometry and
photometric system we use (Colless \etal\ 1993) and the methods we
apply to correct for seeing in measuring the galaxies' light profiles
(Saglia \etal\ 1993b), an important effect for galaxies at greater
distances. Other aspects of this project and some preliminary results
have been reported in Wegner \etal\ (1991), Davies \etal\ (1993) and
Baggley \etal\ (1994).

This paper (Paper~I) describes how the galaxy clusters and galaxies
belonging to those clusters were selected for this project. Future
papers in this series will detail the spectroscopic and photometric
data we have obtained, describe the methods used to analyze the
luminosity profiles of the galaxies, examine the intrinsic properties
of the galaxies and their dependence on environment, derive an optimal
distance estimator, and discuss the peculiar motions of the galaxies
and clusters and their significance for models of the large-scale
structure of the universe.

In \S2 of this paper we describe the selection of the regions and
clusters used in this study, setting them in context with the
surrounding large-scale structures. \S3 gives the procedures used to
select candidate elliptical galaxies in each cluster, and gives the
master list of basic information on the galaxies in the study. The
selection functions for the galaxies in each cluster are quantified in
\S4, and the conclusions we draw from this analysis are given in \S5.

\section{Selection of the Cluster Sample}

We wanted to probe the peculiar velocity field of the galaxies to
greater distances than had been sampled in the 7S study in order to
discover whether the motions found locally within 4000\kms\ are
typical of elsewhere in the universe. In order to achieve this, we
chose two regions of similar size at sufficient distance from each
other and the main parts of the local supercluster that their peculiar
motions should be largely independent of the mass concentrations
producing bulk motions in the local volume. Choosing directions
perpendicular to the Supergalactic plane ensures the maximum
separation between our two regions, and helps avoid possible confusion
with distant parts of the local supercluster.

The depth of the sample is dictated by the choice of the distance
indicator. For example, if a distance indicator has errors of about
20\% for individual objects, then at around 10000\kms\ motions less
than 1000\kms\ can only be detected by averaging over several
galaxies. The strong clustering of elliptical galaxies allows the
selection of sets of galaxies at the same distance, so that the
fundamental plane for elliptical galaxies is a natural choice for the
distance estimator. We have chosen our regions to lie in the range
$cz$=6000--15000\kms: from about twice the outer limit of the 7S
sample to about the practical upper limit of the distance indicator.

An {\em all-sky} survey at this depth would have been considerably
more difficult. As it was, about 350 galaxies had to be observed in
each region to obtain sufficient sampling. Moreover an all-sky survey
is not neccesary to achieve our goal of comparing the bulk motions of
other regions with the local motions within a distance of about 6000 \kms. 
We expect that the geometry of the sample is well-suited for picking out
specific components of the bulk flow if we compare \eg\ with Lauer \& 
Postman, though not as sensitive to
all other directions (Cf. Kaiser 1988; Feldman \&
Watkins 1995; Watkins \& Feldman 1996).
 
These considerations led us to look for regions that were rich in
clusters (so that they could be well sampled) and which lie out of the
Supergalactic plane at distances between 6000\kms\ and 15000\kms. The
selection of suitable regions and clusters (by which we mean
elliptical-rich galaxy associations ranging from Abell clusters to
poor groups) was accomplished in two steps.

(1) {\it Selection of the regions.} Our cluster sample is based on the
Abell (1958) catalog and Jackson's (1982) unpublished list of
elliptical-rich groups and clusters. To select suitable regions we
compiled a list of all the Abell and Jackson clusters with redshifts
in the range $cz$=6000--15000\kms\ as given by Struble \& Rood (1987)
and Jackson (1982).  Examining the distribution of these clusters on
the sky led us to choose two regions which we will refer to as HCB
(Hercules--Corona Borealis) and PPC (Perseus--Pisces--Cetus), although
they do {\em not} correspond precisely to the supercluster complexes
with similar names identified by Tully \& Fisher (1987).  HCB is
bounded by $\alpha = 13^h$ to $19^h$ and 
$90\deg > \delta > -21\deg$ and
PPC by $\alpha = 21^h$ to $06^h$ and $90\deg > \delta > -27\deg$, in both
cases excluding the region with $|b|$$< 10\deg$. The declination
limit in HCB is the southern limit of Jackson's catalog; in PPC we
extended this limit to incorporate some more southerly clusters at
$\alpha \approx 5^h$. 

Subsequent examination of the redshift
distributions in each cluster showed that many have fore- or
background galaxies or groups superposed on them. The 
problems caused by such contamination have been
discussed in the literature (\eg\ Primack \etal\ 1991) and will be
dealt with in the context of our sample 
in a subsequent paper. Here we note that it
complicates the problem of selecting a volume-limited sample of
clusters. In particular, we find that the redshift of an Abell cluster
given in the literature is sometimes that of a bright foreground galaxy
while the true cluster is more distant (see below).

(2) {\it Inspection of the Sky Survey plates.} We next examined glass
copies of all the Palomar Observatory Sky Survey (POSS) E plates in
these two areas and the J plates from the SERC Sky Survey in the
south, identifying the Abell and Jackson aggregates. In addition we
searched the plates to find all elliptical galaxies in each region
with diameters (or redshifts if they appeared in the Huchra redshift
catalog) which indicated that they were likely to be at roughly the
same distance as the Struble \& Rood (1987) and Jackson (1982)
clusters on the plate. This led us to identify some aggregates not
previously cataloged, and in a few cases led us to follow the galaxy
distributions onto neighboring Sky Survey plates. Only those
aggregates which contained at least three ellipticals, as judged by
examining the plates, were retained for further
study.  The KPNO photographic lab then made enlargements of all the
cluster fields found on the plates in order to provide a standard set
of images which would allow us to select galaxies in a more
reproducible way than is possible from the glass Sky Survey
plates. The selection of the galaxy sample is described in detail in
\S3.

Table~1 lists the clusters in our sample. Columns 1-9 give 
respectively the cluster ID number
(CID), $N$, the number of sample galaxies selected in the cluster,
the R.A.\ and Dec.\ (J2000), the Galactic coordinates $(l,b)$,
the number of sample galaxies selected in the cluster $N$, the median
redshift $cz_{EFAR}$ in \kms\ (from the cluster members in our
sample), the redshift from the literature $cz_{lit}$ (obtained using
NED\footnote{NED, the NASA/IPAC extragalactic database, is operated
for NASA by the Jet Propulsion Laboratory at Caltech.}), and the name
of the cluster. The cluster positions are in fact those of the `A'
galaxy in the cluster (normally the brightest; see \S3), and so are
not necessarily coincident with the Abell catalog positions. The table
includes the 84 program clusters we initially selected (CID=1--84) and
the Coma cluster (CID=90), our primary reference cluster. 

For the 40 cases where there is a redshift in the literature, Fig.1
shows the distribution of differences between the median cluster
redshifts we obtain and the literature values (the notes to Table~1
indicate the few cases where we prefer another literature redshift to
the one adopted by NED). The scatter of 312\kms\ 
shown in Fig. 1 is consistent with
the 200--400\kms\ errors expected when estimating the mean redshift of
clusters with line-of-sight velocity dispersions in the range
500--1000\kms\ from 6 galaxy redshifts (the median 
number of usable objects in our sample).
The mean difference of $-49$\kms\ is consistent with the standard
error in the mean expected from the observed scatter.
Better estimates of the individual errors of the cluster 
$cz_{EFAR}$ in Table~1 first require the assignment of
membership to the clusters. That can only be finalized using both redshift and
distance information because of the fore and background objects. That
information will be presented in subsequent papers dealing with the
spectroscopy (Wegner \etal\ 1996) and the photometry (Saglia \etal\ 1996).

The cluster names given in Table~1 are either the Abell number (\eg\
A2052) from the catalogs of Abell (1958) and Abell \etal\ (1989),
Jackson numbers (\eg\ J17) from Jackson (1982), or P-numbers which
combine the number of the Sky Survey field on which the cluster was
found with a sequence number (\eg\ P777-1, P777-2 \etc). Some clusters
were split into suspected components (there is a J34/35 as well as J34
and J35, a A533-1 as well as A533, a A2162-N and a A2162-S, \etc).
Some clusters have alternate names: A85 is also J29, A2152 is also
J19, our A2162-S is A2162, and of course Coma is A1656. A few clusters
were misnamed in Colless \etal\ (1993): A85, A147, A1983, A1991, A2152
and P522-1 were called J29, A150, A1983-1, A1983, J19 and A2506 in
that paper; the names are given correctly in Table~1, and the galaxies
in these clusters are also renamed in the list of sample galaxies (see
\S3 and Table~3 below).

The locations of the survey regions and the 84 program clusters are
shown in Fig.2. The projection in Galactic coordinates is shown for
three redshift shells, with the middle shell corresponding to the
nominal redshift range of our cluster sample, $cz$=6000--15000\kms. In
order to illustrate the level of completeness in our sample, the
figure also shows the positions of all Abell clusters (Abell \etal\
1989; excluding supplementary clusters) with measured redshifts
(extracted from NED in May 1995). Fig.2 also shows the direction of
the Local Group motion with respect to the cosmic microwave background
(Smoot \etal\ 1992) and with respect to the reference frame of Abell
clusters within 15000\kms\ (Lauer \& Postman 1994; Colless 1995). The
HCB and PPC regions are well away from the CMB dipole direction, while
the Lauer \& Postman dipole lies towards the edge of PPC.

A summary of the numbers of Abell ($N_A$), Jackson ($N_J$) and
supplementary P-numbered ($N_P$) clusters and their sum ($N_S$)
as a function of redshift range is
given in Table~2 and 
shows that we were very successful in choosing
clusters in the nominal redshift range 6000--15000\kms. Only 11 of the
84 program clusters are outside this range, and only 2 lie outside the
range 4000--17000\kms. These two clusters are A419 (\#23,
$cz$=20329\kms) and A2148 (\#56; $cz$=26322\kms). They
were selected on the basis of redshifts from Struble \& Rood's
compilation (12180\kms\ and 13250\kms\ respectively) which proved
erroneous. Note that Coma is not a program cluster---although it has
an appropriate redshift, it lies just outside the survey region (it is
the Abell cluster nearest the NGP in Fig.2) and the galaxies in it
were not selected in the same way (see below).

In the nominal redshift range $cz$=6000--15000\kms\ there are a total
of 32 distinct Abell clusters in our sample, 12 in HCB and 20 in
PPC. (The total of 37 in Table~2 arises because we split 5 Abell
clusters into two components due to apparent substructure: A533, A548,
A2063, A2162, A2593.) NED lists a total of 50 Abell clusters in the
survey region over this same redshift range, 15 in HCB and 35 in PPC
(Abell \etal\ 1989, excluding supplementaries). However scrutiny of
the literature references and comparison with the Sky Survey plates
provide strong evidence that the NED redshifts for a significant
number of the Abell clusters which apparently should have been in our
sample are incorrect, mostly belonging to foreground objects (see Appendix~A
for details).

Excluding these clusters as very probably outside our sample redshift
range, we find there are in fact 13 Abell clusters in the HCB volume
and 27 in the PPC volume, so that our samples of Abell clusters are
12/13=92\% complete in HCB and 20/27=74\% complete in PPC. In fact 4
of the clusters missed from our PPC sample are from Abell \etal\
(1989), which was not available when we were selecting our sample;
excluding these our completeness in PPC is 20/23=87\%. The selection
of our cluster sample will need to be accounted for in interpreting
the results obtained on the velocity field.

The location of our survey volumes with respect to the major
large-scale structures is illustrated in Fig.3, which shows the
program clusters relative to the superclusters identified by
Einasto \etal\ (1994). These are very similar to the
superclusters given in the earlier supercluster catalogues of
Bahcall \& Soneira (1984), Batuski \& Burns (1985), Tully \& Fisher
(1987), Tully \etal\ (1992) and Zucca \etal\ (1993), differing only in
the effective density threshold used to define the superclusters and
in having more Abell clusters with redshifts to work with. The names
of the superclusters are those given in Einasto \etal\ (following
earlier authors) except for Pisces~A and Pisces~B, which we have
supplied for Einasto \etal's superclusters 16 and 17. The volume we
call HCB is centered on the Hercules supercluster at 10000\kms, and
reaches towards (but does not encompass) the Corona Borealis and
Bootes superclusters at around 20000\kms. It does not include any
other superclusters identified by Einasto \etal\ The PPC volume
includes the Perseus-Pegasus~A, Pisces~A and Lepus superclusters at
distances around 12000\kms and has an outer boundary that does not
quite include the Pisces~B, Pisces-Cetus and Horologium-Reticulum
superclusters at around 18000\kms.

\section{Selection of the Galaxy Sample} 

The search for suitable ellipticals in each of the selected clusters
was carried out using the high-quality enlargements of the relevant
regions of Sky Survey glass copies described above. These 
enlargement prints greatly aided the uniform selection
of our galaxy sample and the quantification of the selection
criteria. Galaxies were selected entirely by their morphology and size
on the enlargments. As redshifts were unavailable for many of the
galaxies in our program clusters we used the ellipticals with known
redshifts as a guide in finding other galaxies with elliptical
morphologies and similar apparent sizes. We erred on the side of
including some objects with disks rather than exclude possible
ellipticals. In order not to bias the selection, we did not identify
known galaxies, but chose objects solely on the basis of
their appearance on the Sky Survey enlargement prints.

From our previous experience (Faber \etal\ 1989), high-quality
photographic enlargements of the Sky Survey glass copies can be used
to make quantifiable selection of elliptical galaxies. 
Thus an initial survey of the enlargements was conducted
picking out suitably-sized objects that could possibly be E or S0
galaxies. As many galaxy images are saturated on the Sky Survey
plates, we realized at the outset that this selection procedure also
yields spiral galaxies which would have to be weeded out with further
imaging. However we decided it was preferable to use an inclusive
procedure in order to make the selection criteria more readily
quantifiable, and to bear the overhead of the extra subsequent imaging
needed to make final morphological classifications.

All galaxies identified as possible E or S0 galaxies by the initial
selection process were given capital letter designations within each
cluster (\eg\ A119\_A, J3\_C, P777-3\_B). This yielded 598 E/S0 galaxy
candidates. A second pass through the selection process added 145
additional candidates (generally fainter, smaller galaxies), which are
given numerical designations (\eg\ A119\_2, J3\_1) to distinguish them
from the first set. Altogether we selected a total of 743 E/S0 galaxy
candidates in the program clusters. The final list of galaxies useful as
distance indicators was further refined after spectroscopy and spectra were
obtained.

In order to have a calibration and comparison sample, we also chose 52
well-studied galaxies in Coma, Virgo and the field.  These galaxies
were {\em not} selected in the same way as the program galaxies, but
were drawn from samples studied by previous workers investigating the
\Dnsigma\ and fundamental plane relations. Our primary comparison
sample of 32 Coma cluster galaxies are designated by their Dressler
(1984) numbers (\eg\ COMA\_124 is D124). The 7 Virgo cluster and 13
field galaxies are called by their NGC names (\eg\ N4486 is NGC~4486).

Table~3 gives the basic data on all 795 objects observed in the EFAR
project, listing the Galaxy Identification Number (GIN), the galaxy
name, its position (J2000), the R-band extinction $A_R$
derived as described below, the galactic
coordinates $(l,b)$, the log of the photographic diameter $D_W$ in
arcsec (see \S4), and some comments, which include other names for the
galaxies (mainly NGC/IC names). The GIN is the unique identifier for
each galaxy: the 743 program objects are assigned numbers 1 to 742 and
901 (P777-2\_2). The 52 calibration galaxies are assigned numbers 750
to 801. Note that there has been some re-naming of the galaxies
compared to Colless \etal\ (1993) due to the incorrect cluster names
used in that paper. Only the cluster part of the name has been changed
(following \S2 and the notes to Table~1); the alphanumeric designation
(and the GIN) remains the same (\eg\ A2506\_A has become P522-1\_A).

Of the total of 743 E/S0 galaxy candidates it was subsequently found
that there were 3 duplicated pairs (GINs 53=55, 406=435, 565=576) and
that 4 `galaxies' are actually stars (GINs 123, 131, 133, 191); these
are all noted in the comments column of Table~3. Excluding the
duplicates and stars gives a sample of 736 galaxies which are E/S0
candidates. The distribution of the number of galaxies per program
cluster is shown in Fig.4; the number of galaxies selected per cluster
ranges from 2 to 19, with a median of 8.

Accurate positions were determined for all program galaxies using the
Galaxy Automated Scanning Program (GASP) of the Space Telescope
Science Institute. The positions obtained appear to have an accuracy
of 0.5\arcsec, and have been checked repeatedly during acquisition at
the telescope. The R-band extinctions given in Table~3 were computed
as $A_R=2.4E(B-V)$, with $E(B-V)$ obtained from Burstein \& Heiles
(1982, 1984).

\section{Sample Selection Functions}

It has long been known, starting with Malmquist (1920), Neyman \&
Scott (1959) and others, that the statistical properties of a sample
of objects depends on a number of observational 
factors including the cutoff and completeness of
the sample. In investigations of this kind, there are two levels of selection
with which we must deal: (1) the choice of the clusters (or ``fields") which
we have described above and (2) the selection of
the galaxies themselves. Having dealt with the former in \S2
we now turn to quantifying the latter. 

Incompleteness amongst the small, faint galaxies observed
in a given cluster can lead to a strong selection bias which affects
the estimated distance to the cluster (\cf\ Lynden-Bell \etal\ 1988;
Willick 1994; Strauss \& Willick 1995; Freudling \etal\ 1995). As each
cluster has its own intrinsic distance, richness, and structure, it
will also have its own selection function which depends on the
observational quantity on which the sample is selected. The relevant
quantity in this study is the photographic diameter as measured from
the Sky Survey images.

For a cluster of galaxies, index $j$, found within a fixed solid angle
on the sky, we estimated the selection function of the survey (the
completeness as a function of the selection variable $X$), $S_j(X)$,
by binning the program elliptical galaxies in $k$ fixed intervals
$\Delta X$ and summing the number in each bin. We excluded program
galaxies found to be spirals or otherwise unusable
in our subsequent CCD imaging, as these will not be used for obtaining 
distances and so should not be included in the selection function.  This gives 
the count, $N_{jk}^{obs}$, of observed galaxies with the range of
desired morphological types belonging to cluster $j$
in the interval $X_k-\frac{1}{2}\Delta X \leq X_k <
X_k+\frac{1}{2}\Delta X$. The ratio of this to the true number of galaxies
in this cluster and interval with the range of desired morphological types, 
$N_{jk}^{all}$, yields the selection
function,
\beqn 
S_j(X_k) = N_{jk}^{obs}(X_k)/N_{jk}^{all}(X_k) \, .
\eeqn 
In order to estimate $N_{jk}^{all}$ and thus determine $S_j$, a second
catalog of all galaxies that {\em might} have been in the sample must
be constructed. This means, for each cluster, measuring the
photographic diameters for all ellipticals (spirals were
excluded) as big or bigger than the
smallest galaxy that is included in our sample.

The method for selecting candidate galaxies simply consisted of
choosing, by eye, objects larger than some size which looked like
ellipticals on our high-quality photographic enlargements of the glass
copies of the Sky Survey plates. This selection procedure can
be quantified by measuring optical major-axis diameters,
$D_W$ (in arcsec), for every elliptical-like galaxy in all the cluster
fields (including both the galaxies in the sample and other, mostly
smaller, galaxies in the same fields). These diameters were measured
in a homogeneous fashion by GW off the photographic enlargements of
the Sky Survey glass copies. In all, 2185 diameter measurements were
made; those for the program galaxies are listed in Table~3. These
$D_W$ are a good measure of the true size of the galaxies, as is
illustrated by Fig.5, which, using preliminary estimates of $D_R$ (the
diameter enclosing a mean surface brightness of 20.5~mag~arcsec$^{-2}$ 
in the Kron-Cousins $R$ band), shows the good correlation that exists 
between $\log D_W$ and $\log D_R$. The full details of the derivation of the
$D_R$ values will be given in the Paper III on the photometry (Saglia \etal\
1996).

The selection function for each cluster $j$ was computed from the
ratio of the $\log D_W$ distributions (with $\Delta\log D_W = 0.2$~dex
binning) of the EFAR program galaxies (with spirals omitted),
$N_j^{EFAR}$, and {\em all} galaxies with measured $D_W$, $N_j^{all}$:
\beqn
S_j(\log D_W) = N_j^{EFAR}(\log D_W)/N_j^{all}(\log D_W) \, .
\eeqn
To characterise the selection function $S_j(\log D_W)$ we follow Neyman
\& Scott (1959) and Willick (1994) and adopt the useful form
\beqn
S_j(\log D_W)=0.5\{1+{\rm{erf}}[(\log D_W-\log D_{Wj}^0)/\delta_{Wj}]\} \, , 
\eeqn 
where $\log D_{Wj}^0$ is the midpoint, and $\delta_{Wj}$ the width, of
the cutoff in the selection function. In practice, we fitted this
relation by doing a linear least-squares fit to
\beqn
{\rm erf}^{-1}[2S_j(\log D_W)-1] = (\log D_W-\log D_{Wj}^0)/\delta_{Wj} \, ,
\eeqn
where ${\rm erf}^{-1}$ denotes the inverse error function. For some of
the clusters there were too few points to fit both parameters and we
used the mean width $\langle \delta_W \rangle$=0.24 and determined
only the cut-off $\log D_W^0$. The uncertainty in $\log D_W^0$ is
dominated by the small numbers of galaxies in each cluster, and is
approximately 0.1~dex. Example selection functions for some of the
clusters are shown in Fig.6. Note
that the cutoff in the selection function for Coma is particularly
broad, due to the fact that the galaxies in Coma were not selected in
the same way as the other clusters but were simply garnered from
the literature of previous observations in the cluster.

Our estimates of the selection functions for each cluster
are valid as long as (1) the diameters measured for all
galaxies are as small or smaller than the smallest program galaxy in the
cluster and (2) the two samples are not badly contaminated by
galaxies of the wrong morphological types. 
Fig. 7a shows that this first requirement is satisfied, for the
catalog of all $\log D_W$ diameters (striped histogram)
only begins to show incompleteness about 0.2~dex 
below the cutoff $\log D_W^0$ in the sample of program
galaxies in each cluster. 

The second potential source of error in the
galaxy counts stemming from the difficulty of assigning correct
types to the smallest galaxies could produce a systematic change in the
contamination by spirals with diameter. However this would be compensated
approximately if the fractions of incorrectly selected galaxies are 
similar in the two catalogs. This is shown to be the case in Fig. 7b
which compares the distributions of rejected spirals from the two
samples. Here the two distributions have nearly the same shape and 
their ratio remains nearly constant at $\sim$0.5 near the cutoff of the
selection functions. The final morphological types will be given in the
photometry paper (Saglia \etal\ 1996). Here we are concerned purely with the
initial sample selection; this involved the by-eye morphologies, but not the
detailed CCD-image-based classifications which will be discussed in later
papers.

The combined selection function for the whole sample, correcting for
differences in $\log D_W$ between clusters, is shown in Fig.8. The
selection function corresponding to the mean parameters $\langle \log
D_W^0 \rangle$=1.25 and $\langle \delta_W \rangle$=0.24 (\ie\
cluster-weighted using the subsample of clusters for which we
could directly determine these parameters) is shown as 
the solid curve. Fitting the combined galaxy sample directly yields the
dashed curve which is the
galaxy weighted selection function and it has a
slightly higher cutoff $\log D_W^0$=1.30 and is slightly broader
$\delta_W$=0.30.  Thus the selection functions typically have a cutoff
at $\langle D_W^0 \rangle$=18--20\arcsec\ and drop from 90\% to 10\%
completeness over a range of $1.8 \langle \delta_W
\rangle$=0.44--0.54~dex (\ie\ between 30\arcsec\ and 10\arcsec). Fig.9
shows that the selection is indeed by angular diameter: note the
curves of constant angular size produced by the discreteness of the
$\log D_W^0$ estimates. Over the range from 6000\kms\ to 15000\kms\ a
typical cutoff diameter of 19\arcsec\ corresponds to a metric diameter
increasing from 11~kpc to 28~kpc (for $H_0$=50\,km\,s$^{-1}$\,Mpc$^{-1}$).

The selection regions and the parameters of the selection functions 
for all clusters are given
in Table 4. The CID in the first column correspond to those in Table 1. The
solid angles on the sky that were surveyed for each cluster were measured from
the photographic prints and with the exception of two clusters, the fields are
rectangles running EW and NS. Thus we measured the right ascension and
declination of the center of each rectangle (not the same
as the cluster centers in Table 1), given in columns 2 and 3, and the
total length and height, $\Delta \alpha$ and $\Delta \delta$ of the sides of
these rectangles in units of arc minutes in columns 4 and 5. The
selection function parameters $N_W$ (the total number of $D_W$ diameters 
measured in each field), $\log
D_W$, and $\delta_W$ are listed in columns 6, 7, and 8.  

In subsequent papers we will apply additional selection criteria (such
as the availability of spectroscopic and photometric observations 
or morphological
criteria based on the CCD imaging) in order to refine the initial
program sample. We will then compute the corresponding new selection
functions using the method described above. In general we would expect
such subsamples to omit more objects with small diameters, leading to
larger values of the cutoff diameter.
This is borne out using a preliminary stringently chosen list of our highest 
quality distance indicators (Saglia \etal\ 1996). With this whittled 
down subsample which is 55\% of the original list, the mean $\log D_{W}^{0}$ 
increases by only 0.1 while $\delta_{W}$ changes insignificantly and shows the
insensitivity of the selection function to large changes in the sample.

A maximum likelihood fitting method is to be used to determine the
galaxy distances, and since this technique fits a probability
distribution to all observables this gives additional explicit and
implicit selection criteria besides $S_j(D_W)$. The final values that we use
will be given in subsequent papers, but
examples of these with preliminary values are as follows: 
the lower cutoff in the velocity dispersion at
$\sigma$$<$140\kms\ due to the resolution of the instruments is an
explicit limit (Wegner \etal\ 1996), while the mean surface
brightness 23.39$\geq\langle SB_e \rangle$$\geq$16.84~mag~arcsec$^{-2}$ and the
smallest effective radius $R_e$$\geq$1.5\arcsec\ we have observed in
our sample are implicit limits (Saglia \etal\ 1996).

\section{Conclusions}

The EFAR project is aimed at measuring the properties and peculiar
motions of elliptical galaxies in clusters selected in two regions at
distances of 6000--15000\kms. The primary goals of the project are:
first, to study the physical properties of a large sample of
elliptical galaxies and derive an optimal distance
estimator; and second, to determine the bulk motions in the two
selected regions in order to establish whether the large-scale
coherent motions seen within 6000\kms\ are typical of other regions of
the universe, thereby tightening the constraints on cosmological
models set by the local velocity field.

There are 84 clusters in our sample, 39 in the region we call
Hercules-Corona Borealis (HCB) and 45 on the opposite side of the sky
in the region we call Perseus-Pisces-Cetus (PPC). Most of these
clusters lie in the redshift range 6000--15000\kms, with 42 being
drawn from the Abell (1958) catalog, 32 from the catalog of Jackson
(1982), and a further 10 supplementary clusters found by us on the Sky
Survey plates. We give the redshifts for all of these clusters and show
where they are located with respect to the superclusters identified by
Einasto \etal\ (1994). We find that our sample of Abell (1958)
clusters in the range 6000-15000\kms\ is 92\% complete in HCB and 87\%
complete in PPC.

In these clusters we have selected 736 candidate elliptical galaxies
based on their morphology and apparent size on enlargements of Sky
Survey glass copies. Our master list of EFAR galaxies gives the
fundamental data, including accurate new positions, for these program
galaxies and for 52 calibration galaxies in Coma, Virgo and the nearby
field. In order to quantify our selection criteria we have measured
visual diameters from the Sky Survey enlargements for all the program
galaxies (and for the calibration galaxies in Coma) and for a large
number of other galaxies in these clusters---a total of 2185
objects. The visual diameters of the program galaxies turn out to
correlate with preliminary estimates of the 
the photometric sizes established by subsequent
CCD imaging. For each cluster we are therefore able to characterise
the selection criterion for the sample galaxies in terms of these
visual diameters. We find that the samples of galaxies in the program
clusters are typically 50\% complete at 18--20\arcsec, and drop from
90\% to 10\% completeness between 30\arcsec\ and 10\arcsec.

Subsequent papers in this series will report the spectroscopic and
photometric observations, estimate distances and peculiar velocities
for the galaxies and clusters, and interpret the implied peculiar
velocity field.

\acknowledgements

GW is grateful to the SERC and Wadham College for a year's stay in
Oxford, and to the Alexander von Humboldt-Stiftung for making possible
a visit to the Ruhr-Universit\"{a}t in Bochum. MMC acknowledges the
support of a Lindemann Fellowship, a DIST Collaborative Research Grant
and an Australian Academy of Science/Royal Society Exchange Program
Fellowship. RPS acknowledges the support by DFG grants SFB 318 and
375. This work was partially supported by NSF Grant AST90-16930 to DB,
AST90-17048 and AST93-47714 to GW, AST90-20864 to RKM, and NASA grant
NAG5-2816 to EB. The entire
collaboration benefitted from NATO Collaborative Research Grant 900159
and from the hospitality and monetary support of Dartmouth College,
Oxford University, the University of Durham and Arizona State
University. Support was also received from PPARC visitors grants to
Oxford and Durham Universities and a PPARC rolling grant: "Extragalactic
Astronomy and Cosmology in Durham 1994-98." 
The KPNO photographic laboratory produced the extensive
set of prints from the Sky Surveys used for measuring galaxy diameters
and as finders, which were provided by RLD and DB. DB measured 
the positions of all program galaxies
using the GASP system; we extend our great appreciation to the STScI
personnel for their help. Josef Wegner compiled the initial sample of
candidate galaxies and provided a list of cross-references which
proved invaluable.

\appendix
\section{Clusters Omitted from the Survey Sample}

As of mid-1995 there were 56 Abell clusters with known redshifts and
sky positions that would nominally have placed them in the survey
sample. The EFAR sample has 34 (61\%) of these clusters. The 22
remaining Abell clusters were excluded for the reasons listed below.

Five clusters had redshifts known at the time of the 1986 search
but were rejected:
 
(1)~A195 - Nominal $z$=0.047. Examination of POSS print shows that the
cluster is much fainter and likely to have $z$$>$0.05.

(2)~A261 - Nominal $z$=0.0467. Examination of POSS prints shows quoted
redshift is that of a foreground elliptical. The real Abell cluster is
much more distant.

(3)~A407 - Nominal $z$=0.047.  Famous ``cD-in-the-making'', seven
galaxies in a common envelope associated with UGC 2489 (cf. Nilson
1973).  We could not get good data on each part of this complex
system.

(4)~A484 - Nominal $z$=0.0386.  Redshift quoted corresponds to a small
radio galaxy near the cluster, but not in it. The real Abell cluster
is of very faint galaxies in background.

(5)~A539 - Nominal $z$=0.0267.  This cluster lies at low Galactic
latitude, has strong differential reddening, and is in the middle of
an emission line nebula.

Ten clusters did not have redshifts known at the time of the 1986
search and subsequent investigation indicates our survey would not
have used them:

(6)~A256 - Nominal $cz$=13379\kms. Examination of POSS prints finds a
foreground E superimposed on a faint, background cluster. Zabludoff
\etal\ (1993) show this region to have galaxies with $cz$ from 6000 to
30000\kms, but no definite clusters except at the highest $cz$.

(7)~A2995 - Nominal $cz$=11332\kms. Examination of POSS prints finds 3
larger galaxies (2 spirals, 1 elliptical) superimposed on faint
background clusters.  Abell (1958) calls this cluster distance class
5, making it too far away for our survey.  Quoted redshift comes from
NED.

(8)~A480 - Nominal $cz$=14180\kms. Struble \& Rood (1991) report this
as an incorrect redshift. We confirm this from examination of the
POSS, as we only see faint cluster and Abell gives it a distance class
of 5.

(9)~S0449 - Nominal $cz$=14390\kms. Redshift from Dalton \etal\
(1994), but assigned probability of 0.05 or less of being
correct. Examination of SRC~J prints indicate galaxies in cluster have
$z$$>$0.1.

(10)~S0471 - Nominal $cz$=12891\kms. Redshift from Dalton \etal\
(1994), but assigned probability of 0.05 or less of being
correct. Examination of SRC~J prints indicate galaxies in cluster have
$z$$>$0.1.

(11)~A3175 - Nominal $cz$=11691\kms. Redshift from Dalton \etal\
(1994), but assigned probability of 0.05 or less of being
correct. Examination of SRC~J prints indicate galaxies in cluster have
$z$$>$0.1.

(12)~A2022B - Nominal $cz$=9144\kms. Examination of POSS shows this to
be a foreground group to a much fainter Abell cluster. The actual
Abell cluster is too distant for our survey.

(13)~A2025 - Nominal $cz$=13550\kms. Examination of POSS shows only
faint galaxies at the Abell cluster position. Quoted redshift probably
of foreground galaxy.

(14)~A2506 - Nominal $cz$=8660\kms. The confusing picture given by
POSS examination is clarified by Zabludoff \etal\ (1993) which shows
the galaxies in this region have a large range in redshift.  Only the
nearest galaxies are used for the quoted redshift. Our survey
originally identified a set of galaxies at this redshift as A2506, but
these are really a degree away from the Abell cluster position, and
are now called the P522-1 group.

(15)~A2592 - Nominal $cz$=13880\kms. Examination of the POSS finds
only faint galaxies. Struble \& Rood (1991) state that this is an
incorrect redshift for this cluster.

Two clusters did not have redshifts at the time of our 1986 search and
would have been of marginal use for this survey.

(16)~A3367 - Nominal $z$=0.0443. Examination of POSS prints shows this
cluster is dominated by a cD but the other galaxies in the cluster are
too faint and too small for our survey. This could be another
foreground galaxy contaminated-cluster.

(17)~A3374 - Nominal $z$=0.0471. Examination of POSS prints shows this
cluster is dominated by a cD. Only faint galaxies are seen around this
cD, too faint for this survey.

Five clusters would have been included in the EFAR survey had we known
about their redshifts in 1986:

(18)~A154A - $cz$=12837\kms.

(19)~A295 - $cz$=12717\kms.

(20)~A536 - $cz$=11931\kms.

(21)~A2881 - $cz$=13280\kms.

(22)~A3223 - $cz$=12981\kms.



\begin{deluxetable}{rrrrr}
\tablewidth{8.5cm}
\tablecaption{Cluster Sample by Type and Redshift}
\tablehead{$cz$ range & $N_A$ & $N_J$ & $N_P$ & $N_S$ }
\startdata
 & \multicolumn{4}{c}{\dotfill\ HCB \dotfill} \\
     $<$6000 &  0 &  0 &  1 &  1 \\
 6000--15000 & 14 & 17 &  4 & 35 \\
15000--20000 &  1 &  1 &  0 &  2 \\
    $>$20000 &  1 &  0 &  0 &  1 \\
       Total & 16 & 18 &  5 & 39 \\
 & \multicolumn{4}{c}{\dotfill\ PPC \dotfill} \\
     $<$6000 &  1 &  0 &  1 &  2 \\
 6000--15000 & 23 & 12 &  3 & 38 \\
15000--20000 &  1 &  2 &  1 &  4 \\
    $>$20000 &  1 &  0 &  0 &  1 \\
       Total & 26 & 14 &  5 & 45 \\
\enddata
\end{deluxetable}%
                                                    

\begin{figure}
\plotone{efar1_fig1.eps}
\caption{The distribution of differences between our median redshifts
for each Abell cluster in our sample and the cluster redshifts given
by NED. The curve is a Gaussian with mean and standard deviation
determined from the data.}
\end{figure}

\begin{figure}
\plotone{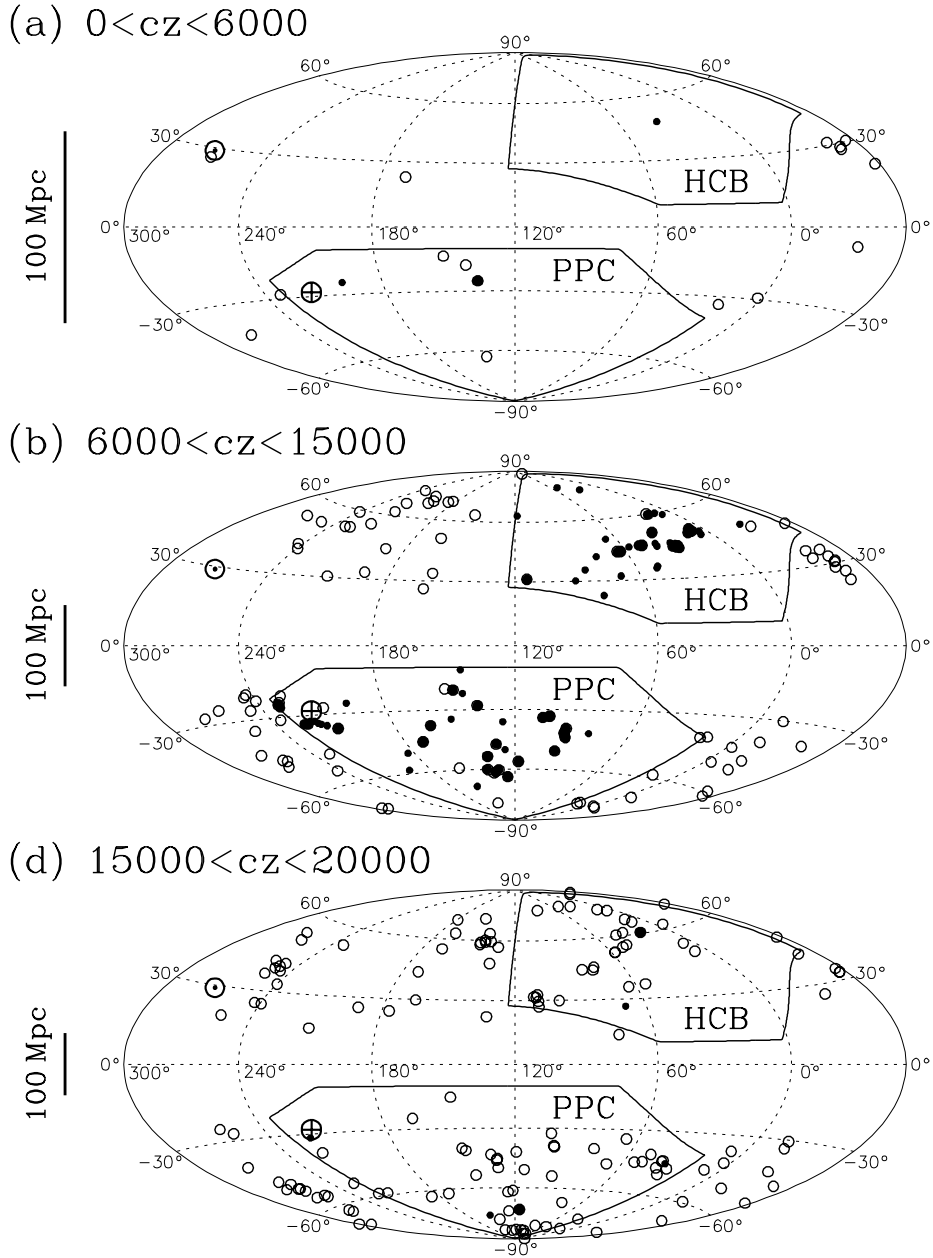}
\caption{The EFAR cluster sample compared to the overall distribution
of Abell clusters. The boundaries of the HCB and PPC regions on the
sky are indicated. Large circles are Abell clusters (filled if they
are in the EFAR sample); small dots are non-Abell clusters in the EFAR
sample. The three panels show the cluster distribution (in an Aitoff
projection of Galactic coordinates) for three redshift ranges:
(a)~$cz$=0--6000\kms, (b)~$cz$=6000--15000\kms,
(c)~$cz$=15000--20000\kms. The scale in \Mpc\ at the upper end of each
redshift range is indicated by the bar at left. Also indicated are the
direction of the Local Group motion with respect to the CMB ($\odot$)
and with respect to the Lauer-Postman 
dipole for the Abell clusters within 15000\kms\ ($\oplus$).}
\end{figure}

\begin{figure}
\plotone{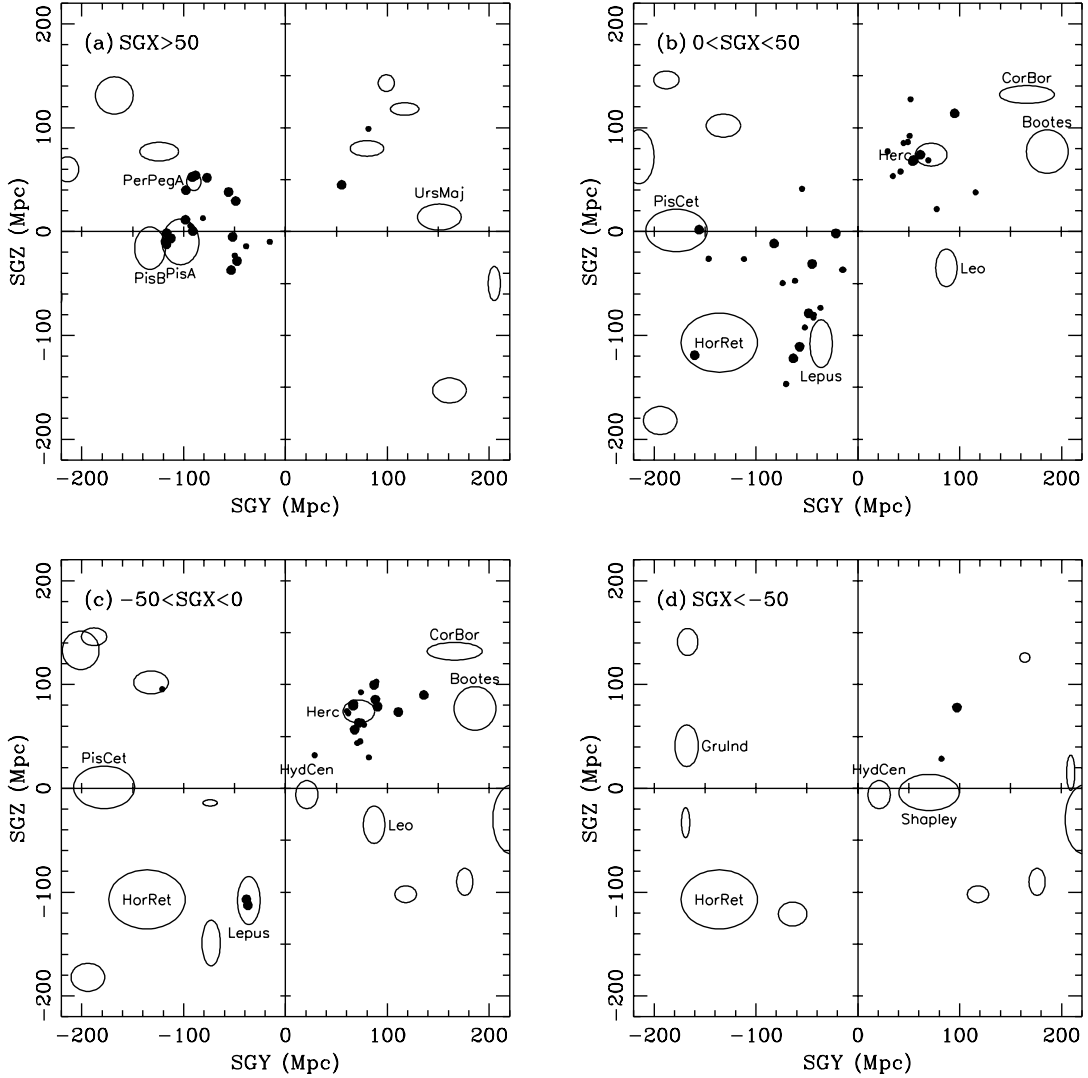}
\caption{The EFAR cluster sample in relation to the surrounding
large-scale structures. Large dots are Abell clusters and small dots
are non-Abell clusters. The ellipses are the superclusters identified
by Einasto \etal\ (1994), with the relevant ones named. The clusters
and superclusters are shown projected onto the supergalactic SGY--SGZ
plane, with the four panels corresponding to four slices in SGX:
(a)~SGX$>$50\Mpc, (b)~SGX=0--50\Mpc, (c)~SGX=$-$50--0\Mpc, and
(d)~SGX$<$$-$50\Mpc.
The size of each supercluster given by Einasto \etal (1994) is represented by
the axes of the ellipses and thus should roughly correspond to the extent of
supercluster.
}
\end{figure}

\begin{figure}
\plotone{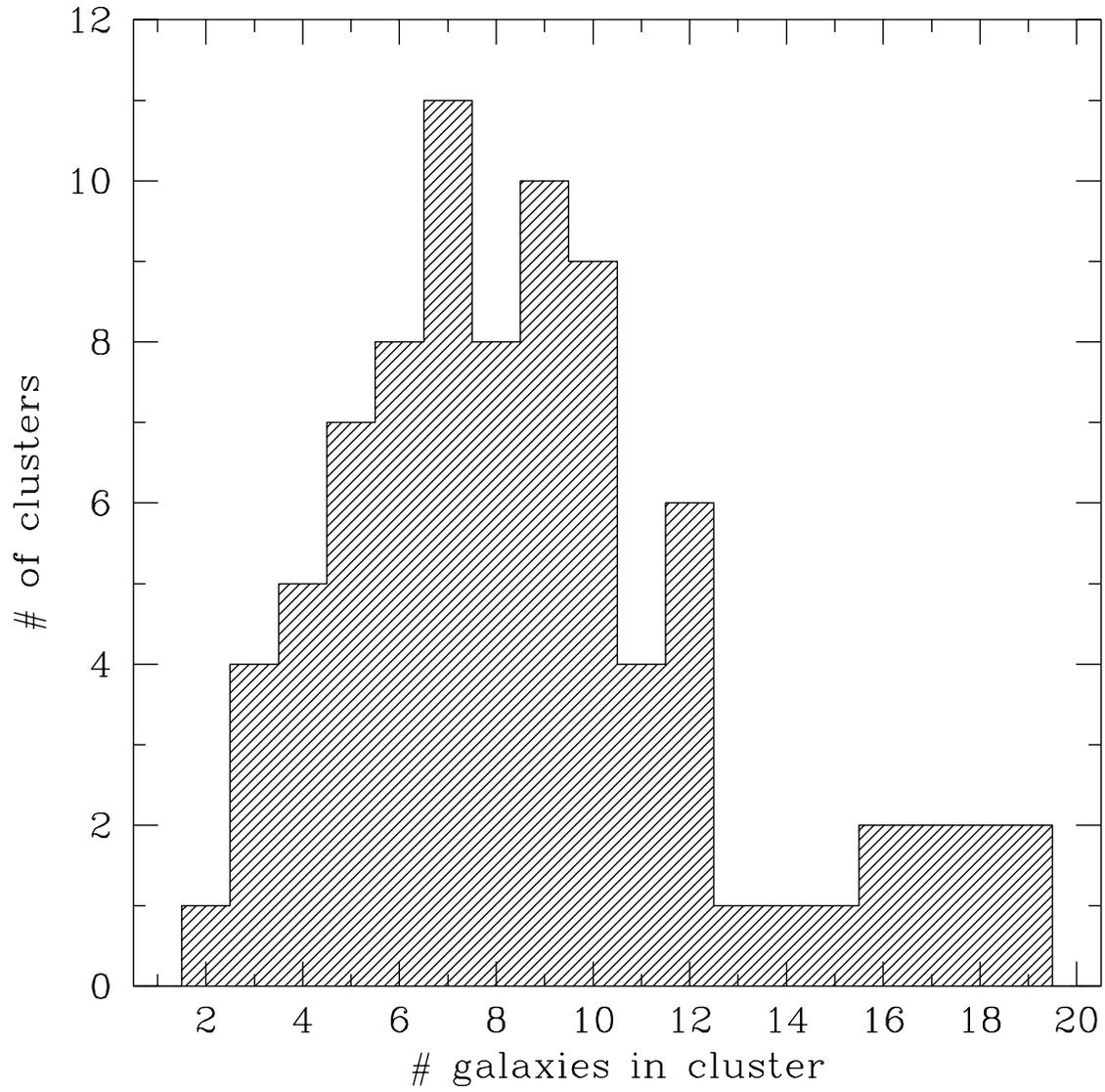}
\caption{The distribution of the number of galaxies selected in each
program cluster.}
\end{figure}

\begin{figure}
\plotone{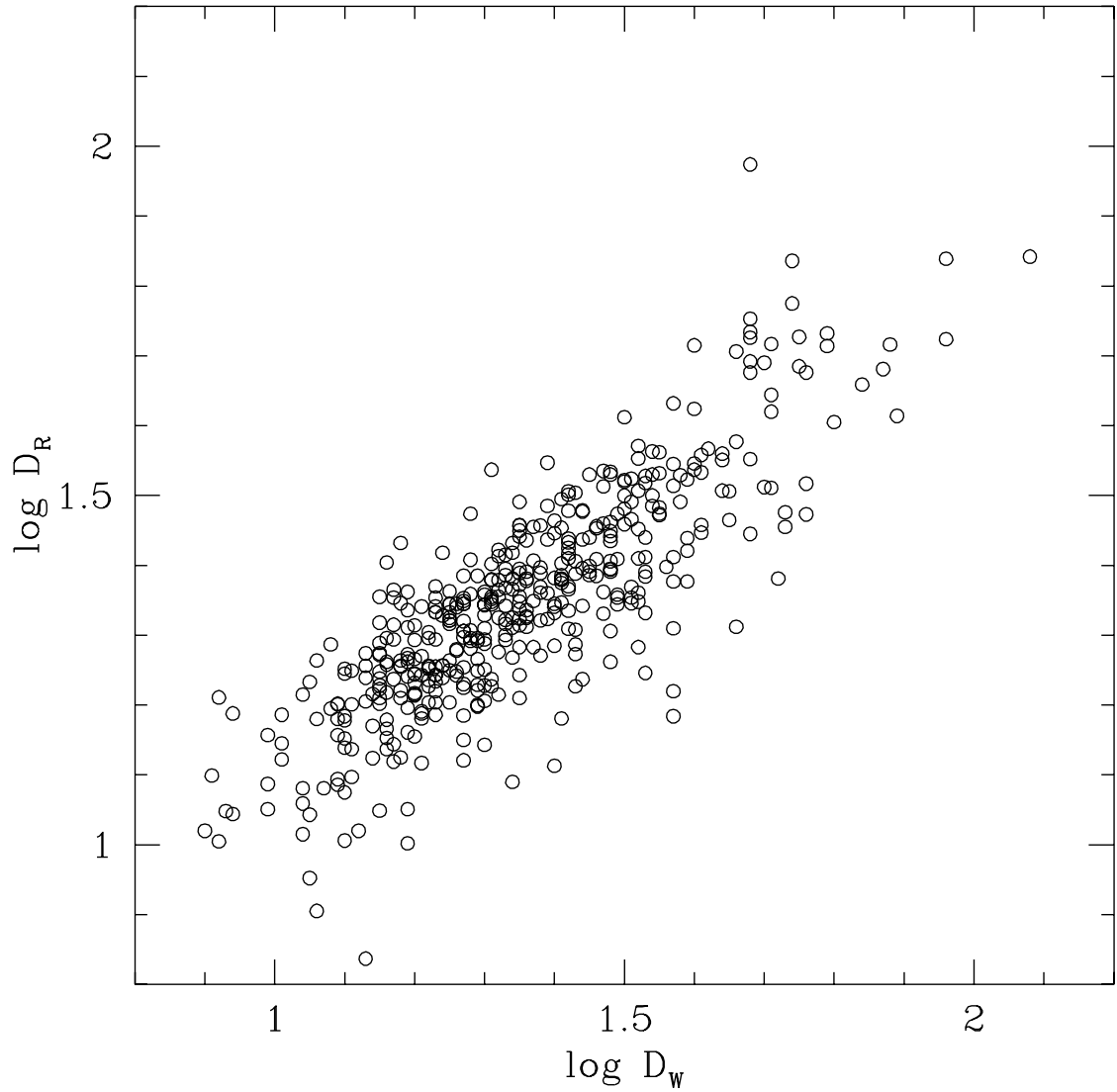}
\caption{The relation between $\log D_W$ (diameters measured by hand)
and $\log D_R$ (photometric diameter at fixed surface brightness) for the whole
galaxy sample, without allowing for offsets between clusters due to
differences in extinction \etc}
\end{figure}

\begin{figure}
\plotone{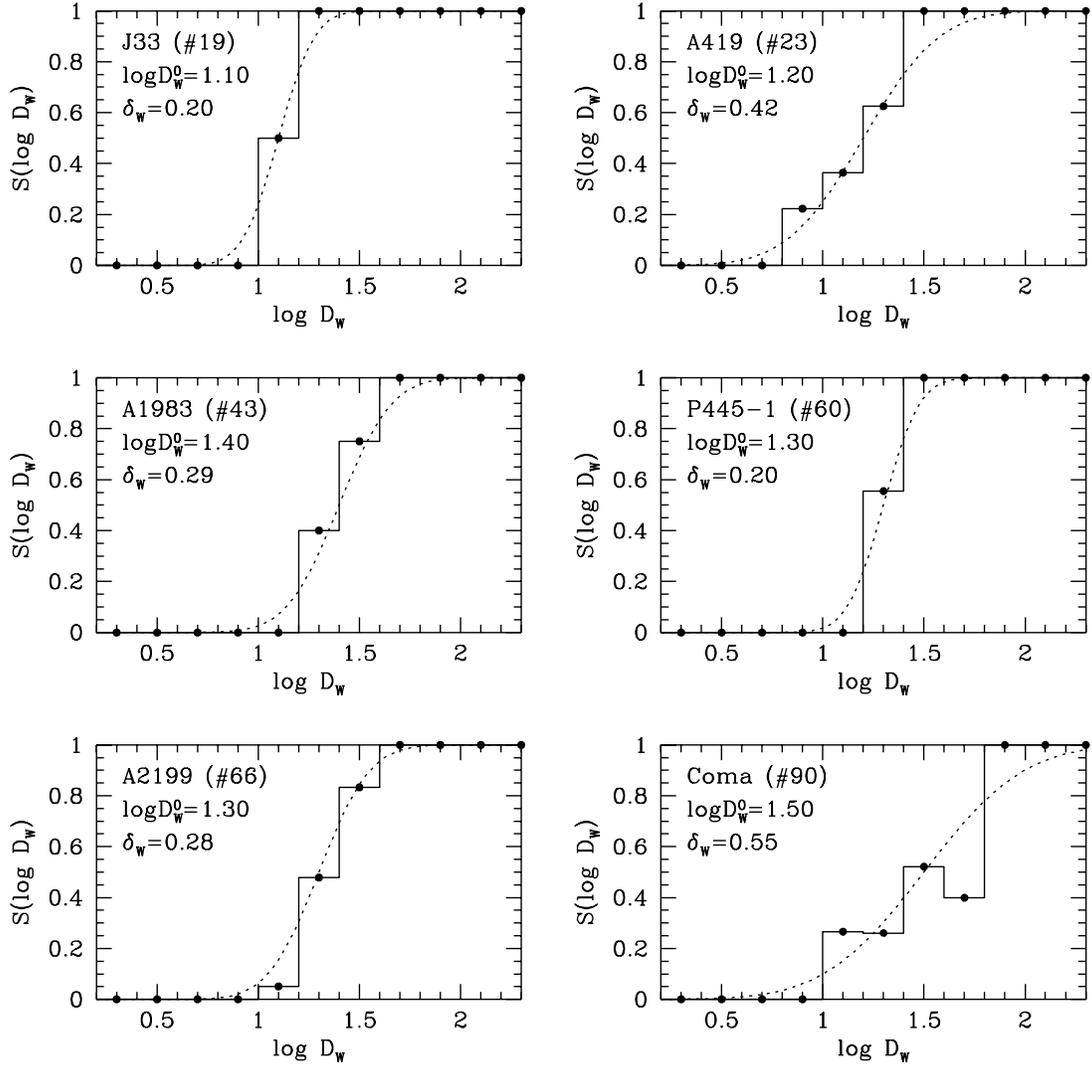}
\caption{Examples of the selection functions for some individual galaxy
clusters. The dots and histograms show the selection function for elliptical
galaxies measured for the cluster whose name is in the upper left corner of
each panel. The dashed curves give the fit to these data using equation (3)
with the cutoff and slope parameters $D_{W}^{0}$ and $\delta_W$.
}
\end{figure}

\begin{figure}
\plotone{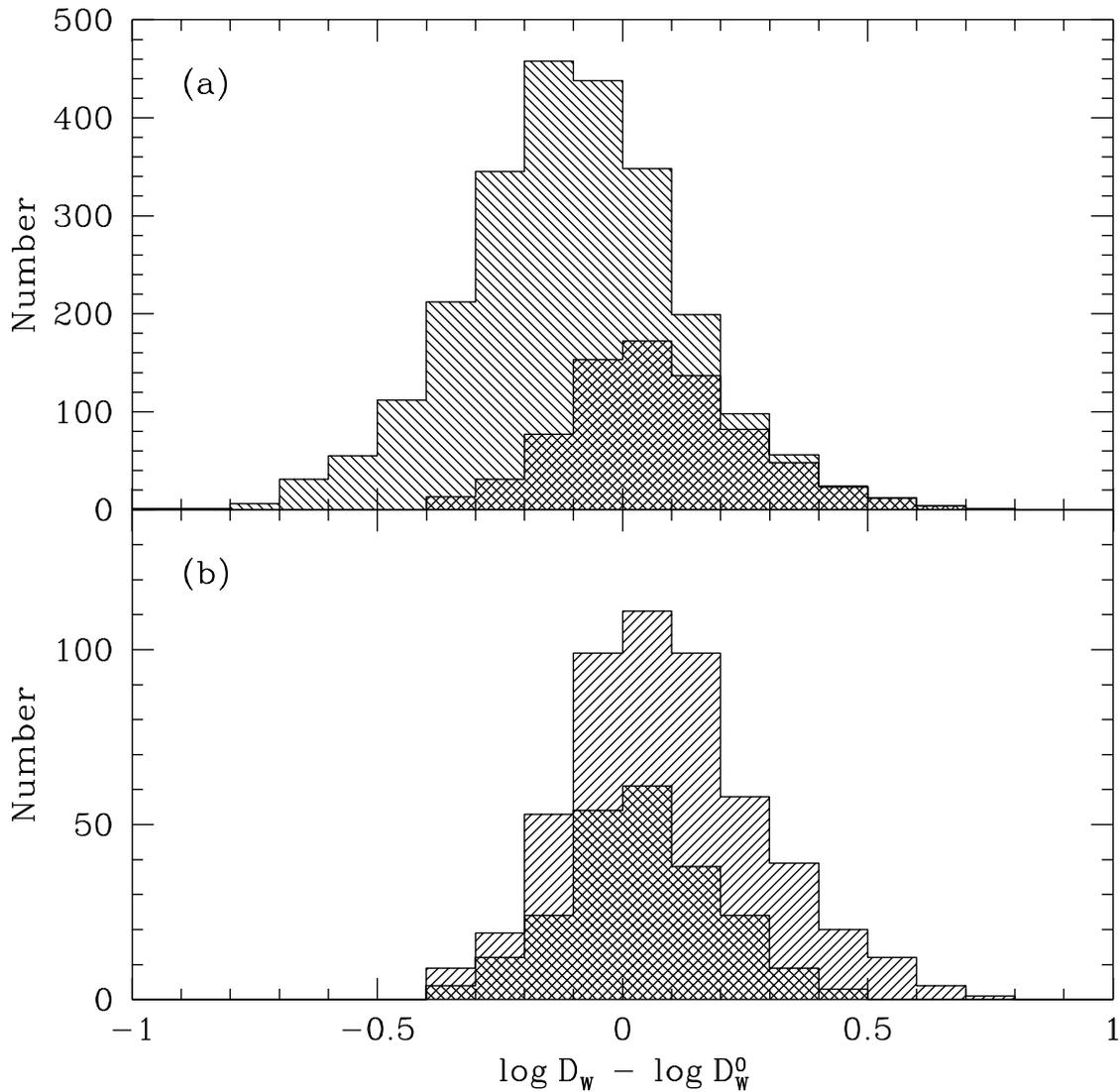}
\caption{(a) The distributions of $\log D_W - \log D_W^0$ for the complete
set of objects (E, S0, and cD galaxies) with $D_W$ diameter measurements 
from the POSS 
(striped histogram) and for subset of the 
the EFAR program galaxies only (hashed histogram). 
(b) The corresponding distributions of spirals rejected from the two
samples.}
\end{figure}

\begin{figure}
\plotone{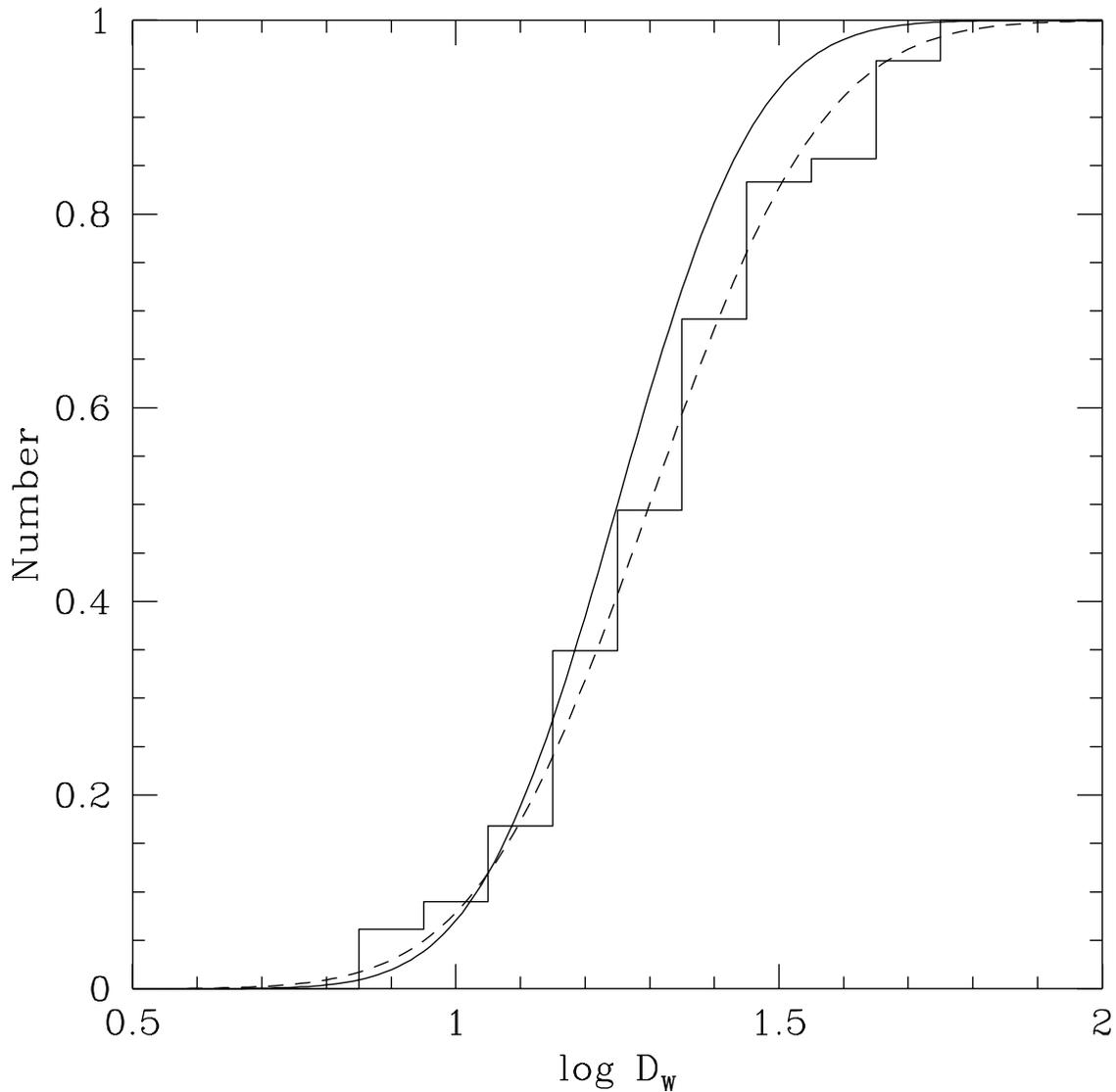}
\caption{The combined selection function for the whole sample, with 
each cluster's selection function shifted to the mean $\log
D_W^0$. The solid curve has $\langle \log D_W^0 \rangle$=1.2 and
$\langle \delta_W \rangle$=0.24, the mean parameters averaged over all
the clusters. The dashed curve has $\log D_W^0$=1.30 and
$\delta_W$=0.30, and is the fit to the combined selection function.}
\end{figure}

\begin{figure}
\plotone{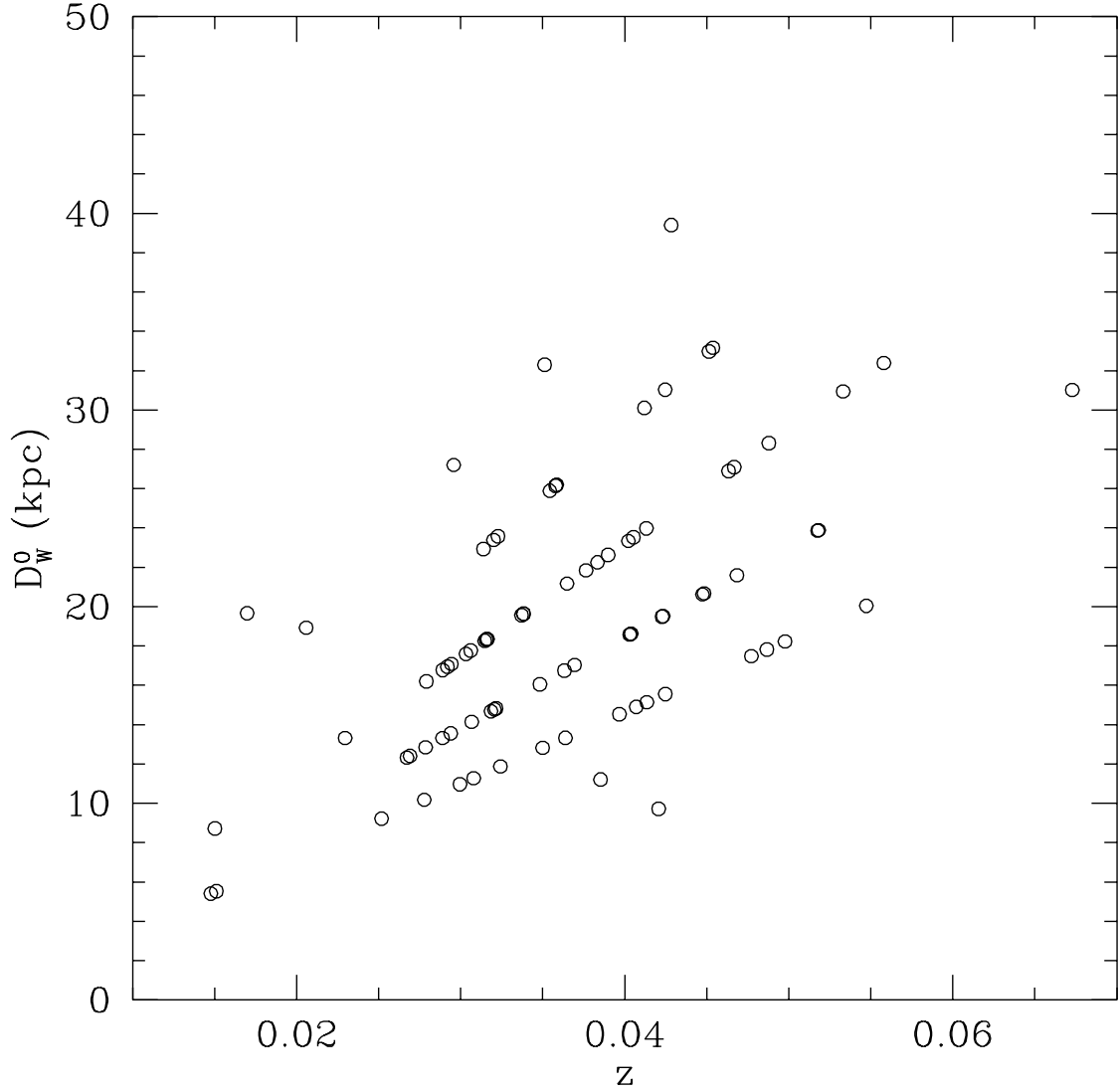}
\caption{The relation between cluster redshift and the cutoff diameter
$D_W^0$ in units of kiloparsecs (for $H_0$=50\,km\,s$^{-1}$\,Mpc$^{-1}$), 
showing that the selection criterion is effectively a constant cut in
angular diameter and thus corresponds to a metric diameter that
increases with redshift.}
\end{figure}

\end{document}